\documentclass[preprint,12pt]{emulateapj}

\usepackage{natbib}

\usepackage{graphicx}
\usepackage[space]{grffile}
\usepackage{latexsym}
\usepackage{amsfonts,amsmath,amssymb}
\usepackage{url}
\usepackage[utf8]{inputenc}
\usepackage{fancyref}
\usepackage{hyperref}
\hypersetup{colorlinks=false,pdfborder={0 0 0},}
\usepackage{textcomp}
\usepackage{longtable}
\usepackage{multirow,booktabs}

\def\icarus{{Icarus}}
\newcommand{\hd}{{HD~106315}\xspace}
\newcommand{\rvjitter}{\ensuremath{6.4\mathrm{~m~s}^{-1}}}
\newcommand{\rvtrend}{\ensuremath{0.3 \pm 0.1\,\mathrm{m~s}^{-1}\mathrm{~d}^{-1}}}

\usepackage{xspace}
\usepackage{amsmath}
\newcommand{\RH}{\ensuremath{R_H}}

\newcommand{\Mearth}{\ensuremath{M_{\oplus}}}
\newcommand{\Rearth}{\ensuremath{R_{\oplus}}}

\newcommand{\Teff}{\ensuremath{T_\mathrm{eff}}}
\newcommand{\vsini}{\ensuremath{v \sin i}}

\shorttitle{At least two bodies orbiting \hd}
\shortauthors{Crossfield et al.}

\begin{document}


\title{Two small, transiting planets and a possible third body orbiting \hd}


\author{
Ian J.\ M.\ Crossfield\altaffilmark{1,2}, 
David R.\ Ciardi\altaffilmark{3}, %
Howard Isaacson\altaffilmark{4}, %
Andrew W.\ Howard\altaffilmark{5}, 
Erik A.\ Petigura\altaffilmark{6,7}, 
Lauren M. Weiss\altaffilmark{8,9},  
Benjamin J.\ Fulton\altaffilmark{10,11,5}, 
Evan Sinukoff\altaffilmark{10,12,5}, %
Joshua E.\ Schlieder\altaffilmark{3,13},%
Dimitri Mawet\altaffilmark{5}, %
Garreth Ruane\altaffilmark{5}, %
Imke de Pater\altaffilmark{4}, %
Katherine de Kleer\altaffilmark{4},%
Ashley G. Davies\altaffilmark{14}, %
Jessie L.\ Christiansen\altaffilmark{3}, 
Courtney D.\ Dressing\altaffilmark{6,2}, %
Lea Hirsch\altaffilmark{4}, 
Bj\"orn Benneke\altaffilmark{6}, 
Justin R.\ Crepp\altaffilmark{15}, %
Molly Kosiarek\altaffilmark{1}, %
John Livingston\altaffilmark{16}, 
Erica Gonzales\altaffilmark{15,1}, %
Charles A.\ Beichman\altaffilmark{3}, %
Heather A.\ Knutson\altaffilmark{6}
}

\altaffiltext{1}{Astronomy and Astrophysics Department, UC Santa Cruz, CA, USA}
\altaffiltext{2}{NASA Sagan Fellow}
\altaffiltext{3}{NASA Exoplanet Science Institute, California Institute of Technology, Pasadena, CA, USA}
\altaffiltext{4}{Astronomy Department, University of California, Berkeley, CA, USA}
\altaffiltext{5}{Astronomy Department, California Institute of Technology, Pasadena, CA, USA}
\altaffiltext{6}{Geological and Planetary Sciences, California Institute of Technology, Pasadena, CA, USA}
\altaffiltext{7}{Hubble Fellow}
\altaffiltext{8}{Institute for Research on Exoplanets, University of Montreal, Canada}
\altaffiltext{9}{Trottier Fellow}
\altaffiltext{10}{Institute for Astronomy, University of Hawai`i at M\={a}noa, Honolulu, HI, USA} 
\altaffiltext{11}{NSF Graduate Research Fellow}
\altaffiltext{12}{NSERC Postgraduate Research Fellow}
\altaffiltext{13}{NASA Goddard Space Flight Center, Greenbelt, MD, USA}
\altaffiltext{14}{NASA Jet Propulsion Laboratory, Pasadena, CA, USA}
\altaffiltext{15}{Department of Physics, University of Notre Dame, 225 Nieuwland Science Hall, Notre Dame, IN, USA}
\altaffiltext{16}{Department of Astronomy, The University of Tokyo, 7-3-1 Bunkyo-ku, Tokyo 113-0033, Japan}

\begin{abstract}
 The masses, atmospheric makeups, spin-orbit alignments, and system
 architectures of extrasolar planets can be best studied when the
 planets orbit bright stars.  We report the discovery of three bodies
 orbiting \hd, a bright ($V = 8.97$~mag) F5 dwarf targeted by our
 {\em K2} survey for transiting exoplanets.  Two small, transiting
 planets have radii of $2.23^{+0.30}_{-0.25}$\,\Rearth\ and
 $3.95^{+0.42}_{-0.39}$\,\Rearth\ and orbital periods of 9.55\,d and
 21.06\,d, respectively. A radial velocity (RV) trend of
 \rvtrend\ indicates the likely presence of a third body orbiting \hd with
 period $\gtrsim160$~d and mass $\gtrsim 45 M_\oplus$. Transits of
 this object would have depths of $\gtrsim$0.1\% and are definitively ruled
 out. Though the star has $\vsini=13.2$~km~s$^{-1}$, it exhibits
 short-timescale RV variability of just \rvjitter, and so is a good
 target for RV measurements of the mass and density of the inner two
 planets and the outer object's orbit and mass. Furthermore, the
 combination of RV noise and moderate \vsini\ makes \hd a valuable
 laboratory for studying the spin-orbit alignment of small planets
 through the Rossiter-McLaughlin effect. Space-based atmospheric
 characterization of the two transiting planets via transit and
 eclipse spectroscopy should also be feasible. This discovery
 demonstrates again the power of {\em K2} to find compelling
 exoplanets worthy of future study.

\end{abstract}


\bibliographystyle{apj}
\section{Introduction}
Planets smaller than Neptune $(R_P \lesssim 4 \Rearth)$ are the most
common type of planet, both in terms of total detections
\citep{coughlin:2016} and intrinsic occurrence
\citep{howard:2010,howard:2012,fressin:2013,petigura:2013a,dressing:2013,dressing:2015}.
Most of these small planets were discovered by NASA's {\em Kepler}
Space Telescope during its prime mission
\citep[2009--2013;][]{borucki:2010}. However, {\em Kepler} surveyed
only 1/$400^\mathrm{th}$ of the sky and thus typically detected planets
orbiting relatively faint stars: extremely useful for demographic
studies, but less so for detailed characterization of planet masses,
spin-orbit alignments, and atmospheric properties.  Small planets orbiting bright host stars are essential
to enable the precise measurements best suited to reveal the
formation, composition, structure, and evolution of these systems.

For planets of a fixed size between 2--4\,\Rearth, the observed masses
span an order of magnitude
\citep{marcy:2014,berta:2015,wolfgang:2016}.  This result indicates
that for a given planet size, many possible bulk compositions are
possible.  Radial velocity (RV)
measurements can determine the mass of a transiting planet and so
constrain its fractional makeup of metal, rock, ice, and gas
(H$_2$/He). Mass and radius measurements alone do not uniquely
determine the bulk makeup of sub-Jovians with radii
$\gtrsim$1.5\Rearth \citep{figueira:2009,rogers:2010,rogers:2011};
further detailed inferences are more difficult when considering that
the atmospheres of these smaller planets may be enhanced in metals by
factors of tens to thousands depending on how the planets formed and
migrated \citep{fortney:2013,moses:2013}.  Atmospheric measurements
are needed to assess the elemental composition of these planets'
atmospheres \citep{crossfield:2015b}, while measurements of the
Rossiter-McLaughlin (RM) effect can constrain planet migration
histories and stellar interiors \citep{winn:2015}. Furthermore, there
is growing interest in comparing all these quantities of planets
orbiting single stars with those of planets orbiting multi-star (or
star-brown dwarf) systems.


As a transit survey {\em K2} lies in the sweet spot between {\em
  Kepler} and {\em TESS} in terms of sky coverage, temporal duration,
photometric precision, and the discovery rate of new candidates
\citep{howell:2014,ricker:2014,sullivan:2015}. Hundreds of planets
have been discovered from the {\em K2} mission, increasing the number
of bright systems ($J=8-12$ mag) known to host small planets
(1--4\,$R_\oplus$) by over 50\% in just its first year
\citep{vanderburg:2016,crossfield:2016}.  Systems such as K2-3,
HD~3167, and HIP~41378 are some of the most interesting of {\em K2}'s
multi-planet discoveries mainly because they are especially good
targets for RV and atmospheric measurements
\citep[][]{crossfield:2015a,vanderburg:2016b,vanderburg:2016c}.

Here, we present the discovery of another new, multi-planet system
around a bright star observed by {\em K2}: two small planets
transiting the F dwarf \hd\ (EPIC~201437844), and a likely RV trend
that would indicate a  third body on a long-period orbit.  The system promises
to be a good target for future RV measurements to explore the system
architecture and planet mass \& spin-orbit alignment, and for future
atmospheric characterization. We describe our discovery, observations,
and derived system properties in Sec.~\ref{sec:obs}, and summarize and
discuss the potential for future observations in
Sec.~\ref{sec:conclusions}.


\section{Observations and Analysis}
\label{sec:obs}

\hd\ was proposed as a {\em K2} target for Campaign 10 (C10) in three
programs: GO-10028 (PI Quarles), GO-10051 (PI Cochran), and by our
team's GO-10077 (PI Howard). The star's basic parameters are listed in
Table~\ref{tab:stellar}. It and other targets in C10 were scheduled to
be observed for the usual $\sim$75~d duration, but during C10's first
six days the spacecraft mispointed by 3.3~pixels (13''). Data acquired
during these first six days is therefore of low quality and so we
discard these early data.  Of the remaining time in C10, a fault with
one of the spacecraft's photometry modules caused an additional 14~d
to be lost before the final seven weeks of C10 observations (see
Fig.~\ref{fig:fits}).

\hspace{-1in}
\begin{deluxetable}{l l l }[bt]
\hspace{-1in}\tabletypesize{\scriptsize}
\tablecaption{  Stellar Parameters of \hd \label{tab:stellar}}
\tablewidth{0pt}
\tablehead{
\colhead{Parameter} & \colhead{Value} & \colhead{Source}
}
\startdata
\multicolumn{3}{l}{\hspace{1cm}\em Identifying information} \\
$\alpha$ R.A. (hh:mm:ss) & 12:13:53.39 & \\
$\delta$ Dec. (dd:mm:ss) & -00:23:36.54 & \\
EPIC ID & 201437844 & \cite{huber:2016} \\
\multicolumn{3}{l}{\hspace{1cm}\em Photometric Properties} \\
B (mag)..........  & 9.402 $\pm$ 0.022 & APASS \\
V (mag) .......... & 8.951 $\pm$ 0.018 & APASS \\
g (mag)..........  &10.14  $\pm$ 0.19  & APASS \\
r (mag) .......... & 9.41  $\pm$ 0.29  & APASS \\
i (mag)........... & 8.848 $\pm$ 0.060 & APASS \\
J (mag)..........  & 8.116 $\pm$ 0.025 & 2MASS\\
H (mag) .........  & 7.962 $\pm$ 0.040 & 2MASS\\
Ks (mag) ........  & 7.853 $\pm$ 0.020 & 2MASS\\
W1 (mag) ........  & 7.794 $\pm$ 0.025 & AllWISE\\
W2 (mag) ........  & 7.850 $\pm$ 0.020 & AllWISE\\
W3 (mag) ........  & 7.839 $\pm$ 0.022 & AllWISE \\
W4 (mag) ........  & 8.168 $\pm$ 0.354 & AllWISE \\
\multicolumn{3}{l}{\hspace{1cm}\em Spectroscopic and Derived Properties} \\
$\mu_{\alpha}$ (mas~yr$^{-1}$) & -1.68 $\pm$ 0.64 & GAIA (\citeyear{gaia:2016}) \\
$\mu_{\delta}$ (mas~yr$^{-1}$) & 11.91 $\pm$ 0.46 & GAIA (\citeyear{gaia:2016}) \\
Distance (pc) & $107.3 \pm 3.9$ & GAIA (\citeyear{gaia:2016})\\
Age (Gyr) & 4$\pm$1 Gyr & HIRES, this paper \\
Spectral Type & F5V & \cite{houk:1999}\\
$[$Fe/H$]$ & -0.24 $\pm$ 0.04 & HIRES; SM\\
\Teff\ (K) & 6290 $\pm$ 60 & HIRES; SM\\
$\log_\mathrm{10} g$ (cgs) & 4.29 $\pm$ 0.07 & HIRES; SM\\
\vsini\ (km~s$^{-1}$) & 13.2 $\pm$ 1.0  &  HIRES; SM\\
$S_{HK}$    & 0.1400 $\pm$ 0.0005 & HIRES\\
$M_*$ ($M_\odot$) &  1.07$\pm$0.03 & HIRES; SM; iso \\
$R_*$ ($R_\odot$) &  1.18$\pm$0.11 & HIRES; SM; iso \\
$L_*$ ($L_\odot$) &  1.95$\pm$0.38 & HIRES; SM; iso \\
$dv / dt$ (m~s$^{-1}$~d$^{-1}$) & \rvtrend & HIRES 
\enddata
\tablenotetext{}{SM: \texttt{SpecMatch} \citep{petigura:2015phd}.
  iso: \texttt{isochrones} \citep{morton:2012}.}
\end{deluxetable}


\subsection{K2 Photometry}
\begin{figure*}[ht!]
  \vspace{-0.1in}
  \begin{center}
\includegraphics[width=7in]{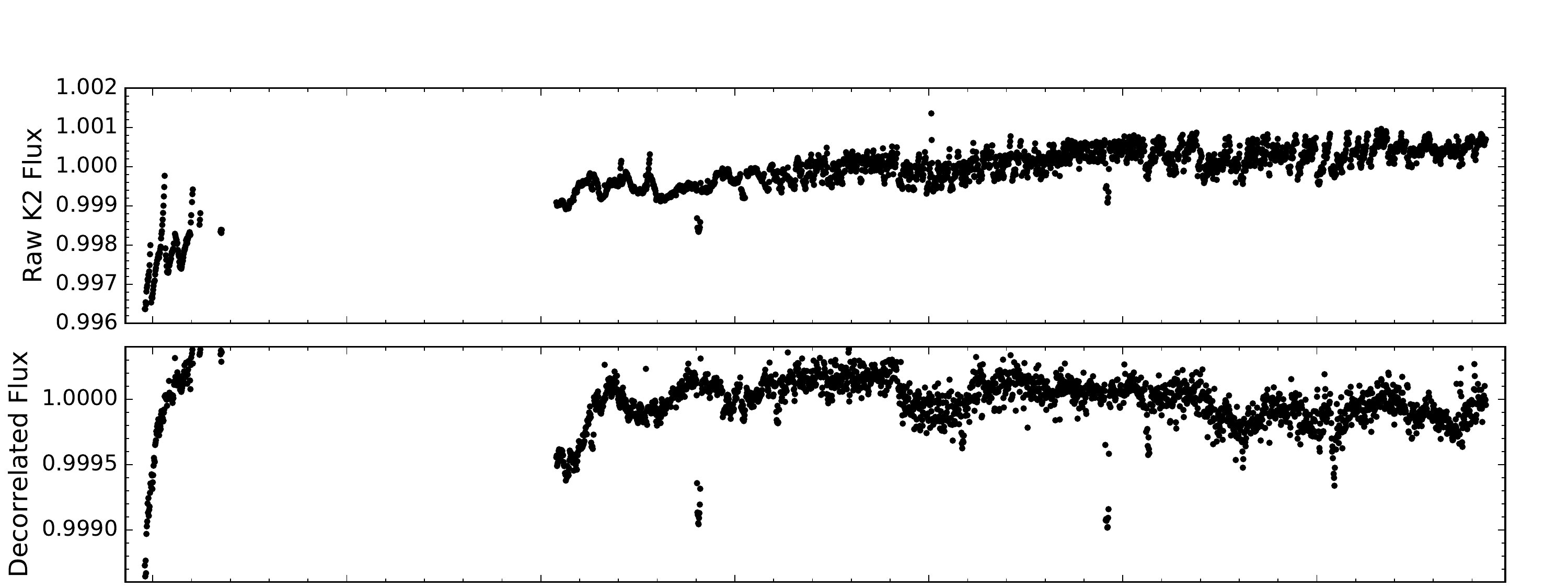}
\includegraphics[width=7in]{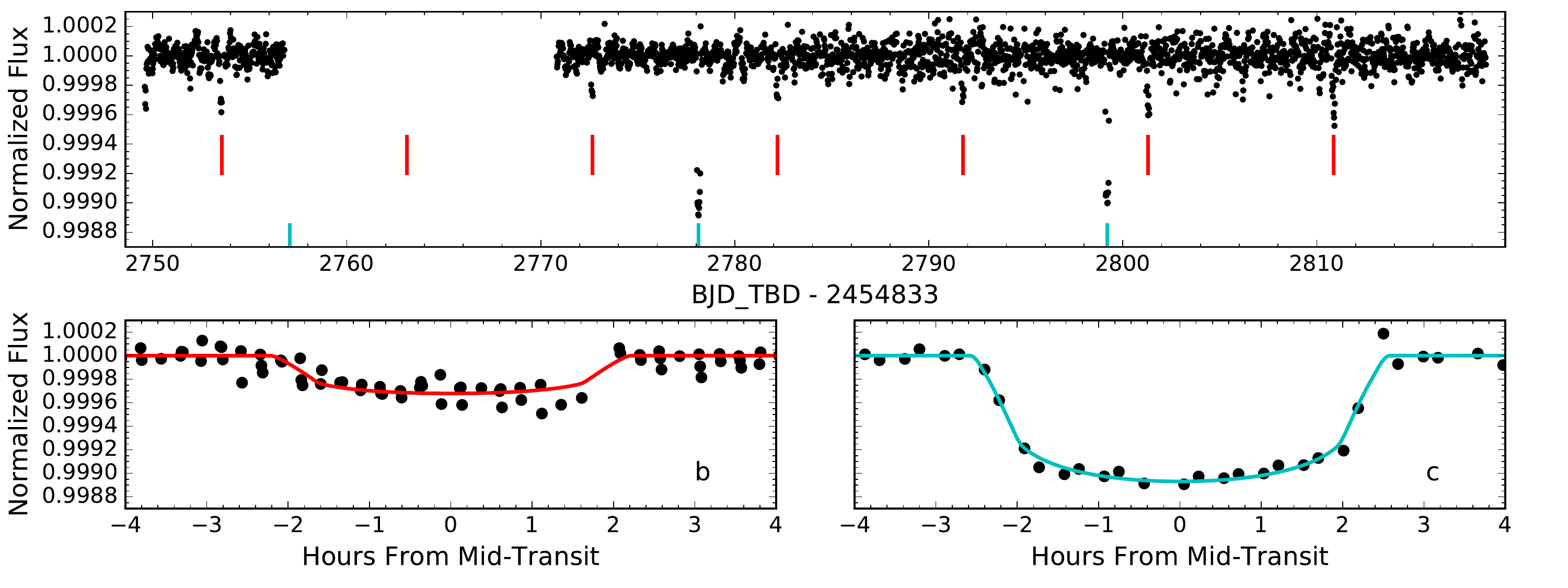}
\caption{\label{fig:fits}From top to bottom: our K2 photometry
  extracted from K2 pixel-level data; the data after decorrelation
  with \texttt{k2phot}; the data after smoothing and detrending, with
  vertical ticks indicating the locations of each planets' transits;
  and at bottom, the phase-folded photometry and best-fit light curves
  for each  transiting planet.}
\end{center}
\end{figure*}

We convert the processed {\em K2} target pixel files from C10 into
light curves and search for transits using the same approach described
in our previous papers
\citep[e.g.,][]{sinukoff:2016,crossfield:2016}. Our light curves at
each step in our analysis are shown in Fig.~\ref{fig:fits}. In brief,
we use the publicly available {\tt k2phot} photometry
code\footnote{\url{https://github.com/petigura/k2phot}} which uses
Gaussian Processes to model out systematics associated with the
$\sim$1~pixel pointing jitter of the spacecraft that occurs over
$\sim$6~hr timescales.  We then use the publicly available
\texttt{TERRA}
algorithm\footnote{\url{https://github.com/petigura/terra}}
\citep{petigura:2013b,petigura:2013a} to search for transit-like
events and manually examine light curves and diagnostic plots for all
plausibly transit-like signals for S/N\,$\ge$\,12.
We discovered a signal with period $P=9.55$\,d in the photometry for
\hd. Inspection of the light curve revealed two deeper transits
separated by 21\,d; a third event was presumably missed during C10's
14\,d data gap (see Fig~\ref{fig:fits}). 

As in our previous work, light-curve fits and an MCMC analysis provide
orbital and system parameters \citep[\texttt{emcee} and
  \texttt{BATMAN};][]{foreman-mackey:2012,kreidberg:2015b}, whose
final distributions are unimodal.  We impose priors on the stellar
limb profile using a quadratic parametrization, with values and
uncertainties derived from \texttt{PyLDTK} \citep{parviainen:2015};
our previous analyses show that this choice does not strongly affect
the system parameters we measure
\citep{crossfield:2016}. Fig.~\ref{fig:fits} shows the resulting
photometry and best-fit models, and Table~\ref{tab:planet} summarizes
the final values and uncertainties.

Several other features are also visible in the intermediate panels of
Fig.~\ref{fig:fits}. First, our data lacks coverage during the transit
egress of \hd b because several {\em K2} thruster-firings occur during
this time.  We did examine {\em K2}'s early (mispointed) observations
of this system, in which we see another transit of \hd b, with
consistent depth (albeit at low S/N). We also see planet b's transits
in the photometry provided by the Kepler Project Office, though 
these data are of lower quality than ours. We see a few low points
that occur together near time index 2750; we attribute these to
uncorrected systematics rather than a transiting object, because we do
not see additional transits at this depth and because these data occur
at the beginning of {\em K2}'s observations (when we see a strong ramp
in the decorrelated data).

We also see several transit-like events in the decorrelated flux panel
of Fig.~\ref{fig:fits}, the most convincing of which occurs at time
index 2790. We fit a transit model to these data and run an MCMC
analysis in which all parameters but $e$ and $\omega$ are
unconstrained, finding a mid-time of $2790.382\pm0.056$ and
$R/R*=0.0133 \pm 0.0012$.  However, the event's profile is asymmetric
(with a much shorter ingress than egress) and we see several other
features with comparable shapes and amplitudes in our data.  Although
an intriguing candidate, we cannot conclude that this signal is
planetary in origin.

\subsection{Ground-based Characterization and \& Validation}

Fortuitously, approximately thirty minutes after identifying \hd as a
set of interesting planet candidates we were able to begin observing
the system using both the Keck/HIRES high-resolution optical
spectrograph \citep{vogt:1994} and the Keck adaptive optics (AO)
system and NIRC2, its near-infrared camera. Below we describe the
acquisition and analysis of these data.

\subsubsection{Keck/HIRES Optical Spectroscopy}

We acquired three HIRES exposures of \hd on UT 2016 Dec 24 to
construct a stellar template for RV analysis and for stellar
characterization. These observations used the B3 decker, had exposure
lengths of roughly 190~s, were acquired in seeing of 1.0--1.1'', and
did not use the instrument's iodine gas cell (used for precise RV
measurements; see below).  We use the \texttt{SpecMatch} algorithm
\citep{petigura:2015phd} to derive stellar properties from our
Keck/HIRES spectrum. The resulting values, shown in
Table~\ref{tab:stellar}, indicate that \hd\ is somewhat larger and
hotter, and rotates more rapidly, than the Sun.
These stellar parameters are generally consistent with, but more
precise than, those derived using broadband photometry and proper
motions only \citep{huber:2016}.

We also use the  Keck/HIRES spectra to search for evidence
of secondary stellar lines, as might be caused by a blended eclipsing
binary \citep{kolbl:2015}. We find no evidence of stellar companions
 down to a sensitivity of 1\% of the brightness of the primary.
Due to the rapid rotation of \hd, we are not sensitive to any
star with a relative velocity within 20 km~s$^{-1}$ of \hd.

The values of \vsini\ and $R_*$ derived above indicate a stellar
rotation period of $\le4.5$~d. After masking out transits and the
first six days of C10 photometry, a Lomb-Scargle periodogram of the
photometry shows a hint of periodicity at $P\approx5.1$ and
$\approx8.5$\,d, with amplitudes of 0.1--0.2\,\%.  The former could be
marginally consistent with a stellar rotation period if the star is
seen nearly equator-on, but would be much more rapid than expected
from gyrochronology \citep{ceillier:2016}. In either case, the low photometric
variability indicates that \hd 's surface is relatively unaffected by
prominent features such as starspots that would modulate the star's
apparent brightness and induce non-planetary radial velocity signals.

Various sources of error, both instrumental and astrophysical, can
mask the Doppler signals from orbiting planets.  This RV ``noise'' is manifest
from multiple physical sources that vary with stellar parameters,
including temperature, surface gravity, and age \citep[see,
  e.g.,][]{howard:2010b}.  For stars cooler than the Sun, rotational
modulation of surface features including faculae and spots is often
the dominant effect
\citep{isaacson:2010,dumusque:2011b,dumusque:2011c,haywood:2014}.
Granulation and acoustic oscillations are also detectable for
magnetically quiet Sun-like stars \citep{dumusque:2011a}.  For stars
hotter and lower gravity than the Sun (such as \hd), surface oscillations can
produce significant false Doppler shifts and high rotational speeds
degrade the quality of the observed spectra, compromising Doppler
precision.  For A--F type stars, the formula of \citep{galland:2005}
predicts the RV scatter in stars observed at high SNR with Elodie and
HARPS, $\sigma_\mathrm{RV} \approx 0.16 \times v \sin i^{1.54}$.  This
formula is accurate at the factor-of-two level and predicts a per-shot RV uncertainty
of 8~m\,s$^{-1}$ for \hd, which is comparable to the \rvjitter\ of
scatter that we typically observe on a given night (see below).
Surface oscillation amplitudes scale as the light-to-mass ratio,
$v_\mathrm{osc} = 0.234(L_\star/M_\star)$ m\,s$^{-1}$
\citep{kjeldsen:1995}.  Although \hd\ is hotter than the Sun,
$L_\star/M_\star$ = 1.8 and surface oscillations do not dominate.

We obtained several epochs of RV observations of \hd\ using Keck/HIRES
with the standard CPS setup: the C2 decker (for all but the first five
RVs, which used the B5 decker), the HIRES iodine cell \citep[used to
  measure precise RVs;][]{marcy:1992}, and exposures of 3--6\,min
(depending on seeing conditions).  These RV measurements are shown in
Table~\ref{tab:rv}.

Although our RV data are not sufficient to robustly measure the masses
of our two transiting planets, we examined our measurements to better
understand the system's RV behavior.  When we subtract each night's
mean from our measurements the RMS of the data drops to \rvjitter,
which is our best estimate for the system's RV noise floor on these
timescales. This noise level is sufficiently low that precise RV
measurements should eventually be able to constrain the masses of the
transiting planets and to characterize the third body's full orbital
properties.



We fit several different models to our RV measurements using
\texttt{radvel}\footnote{\url{https://github.com/California-Planet-Search/radvel}}. We
examine all cases either including or omitting a linear trend; a
sinusoidal planetary signal phased to planet b's orbit; a sinusoidal
planetary signal phased to planet c; and a two-planet model. In all
fits we hold \texttt{radvel}'s ``jitter'' (extra noise) term fixed at
6.4~m~s$^{-1}$ in order to give $\chi^2$ equal to the number of
degrees of freedom for the most complex model considered.
Table~\ref{tab:rvresults} lists the results of these analyses,
including the measured trend and planetary signals (if any) and the
$\chi^2$ and Bayesian Information Criterion (BIC).

It is clear from Table~\ref{tab:rvresults} that the most favored
models all include a linear velocity trend (constant acceleration)
with an amplitude of roughly \rvtrend. The residuals to a trend-only
model have an RMS of 8.7~m~s$^{-1}$, substantially higher than our
night-to-night noise floor reported above; it is therefore likely that
additional coherent RV signals are present above the noise
floor. Indeed, the most favored model includes a trend and RV signals
from both planets, but the planetary semi-amplitudes should be
considered preliminary in light of the sparse data coverage, high
noise levels, and possibility of additional planetary
signals. Nonetheless the best model without a trend is disfavored by
$\Delta$BIC\,=\,16.3, which strongly indicates the presence of the
modeled trend. We also tested models with curvature but find that they
do not improve the BIC.  Below in Sec.~\ref{sec:conclusions} we
discuss the implication of the detected trend.

\begin{deluxetable*}{l l l l}[bt]
\tabletypesize{\scriptsize}
\tablecaption{  Planet Parameters \label{tab:planet}}
\tablewidth{0pt}
\tablehead{
\colhead{Parameter} & \colhead{Units} & \colhead{b} & \colhead{c} 
}
\startdata
   $T_{0}$ & $BJD_{TDB} - 2454833$ & $2772.6521^{+0.0042}_{-0.0045}$ & $2778.1310^{+0.0012}_{-0.0012}$  \\
       $P$ &          d & $9.5521^{+0.0019}_{-0.0018}$ & $21.0576^{+0.0020}_{-0.0019}$  \\
       $i$ &        deg & $88.4^{+1.1}_{-2.1}$ & $89.42^{+0.40}_{-0.67}$  \\
 $R_P/R_*$ &         \% & $1.708^{+0.188}_{-0.083}$ & $3.034^{+0.163}_{-0.067}$  \\
  $T_{14}$ &         hr & $3.96^{+0.17}_{-0.16}$ & $4.693^{+0.078}_{-0.062}$  \\
  $T_{23}$ &         hr & $3.73^{+0.17}_{-0.19}$ & $4.354^{+0.062}_{-0.086}$  \\
   $R_*/a$ &         -- & $0.0599^{+0.0231}_{-0.0065}$ & $0.0299^{+0.0056}_{-0.0016}$  \\
       $b$ &         -- & $0.47^{+0.31}_{-0.32}$ & $0.34^{+0.28}_{-0.23}$  \\
$\rho_{*,circ}$ & g~cm$^{-3}$ & $0.97^{+0.40}_{-0.60}$ & $1.60^{+0.28}_{-0.65}$  \\
       $a$ &         AU & $0.09012^{+0.00083}_{-0.00085}$ & $0.1526^{+0.0014}_{-0.0014}$  \\
     $R_P$ &      \Rearth & $2.23^{+0.30}_{-0.25}$ & $3.95^{+0.42}_{-0.39}$  \\
 $S_{inc}$ &      $S_\oplus$ & $240^{+48}_{-43}$ & $83^{+16}_{-15}$  
\enddata
\end{deluxetable*}

\begin{table}
  \caption{Keck/HIRES Radial Velocities\label{tab:rv}}
  \begin{tabular}{lrr}
\toprule
        HJD &     RV  &    $\sigma_{RV}$$^a$  \\
(UTC) & [m~s$^{-1}$] & [m~s$^{-1}$] \\
        \midrule
 2457746.13805 &   5.7 &  3.9 \\
 2457746.14276 & --0.3 &  3.9 \\
 2457747.06857 &   5.9 &  3.7 \\
 2457747.10475 &   5.5 &  3.6 \\
 2457747.15905 &  17.2 &  3.5 \\
 2457760.09504 &  10.7 &  3.7 \\
 2457760.13026 & --8.8 &  3.7 \\
 2457760.17270 & --2.2 &  3.8 \\
 2457764.01673 &  12.1 &  3.8 \\
 2457764.05201 &   8.8 &  3.6 \\
 2457764.08954 &   5.4 &  3.5 \\
 2457764.09291 &  11.1 &  3.6 \\
 2457764.09626 &  15.3 &  3.6 \\
 2457764.13194 & --0.6 &  3.6 \\
 2457764.17179 &  15.4 &  3.2 \\
 2457765.02290 & --7.3 &  3.2 \\
 2457765.02811 &   1.7 &  3.6 \\
 2457765.03199 &   1.8 &  3.4 \\
 2457765.06751 & --3.4 &  3.4 \\
 2457765.14384 &   1.1 &  3.1 \\
 2457765.15072 & --2.7 &  3.4 \\
 2457765.15814 &   2.1 &  3.2 \\
 2457766.01963 &   4.6 &  3.4 \\
 2457766.05401 & --8.4 &  3.5 \\
 2457766.10269 &--14.6 &  3.4 \\
 2457766.13235 & --7.8 &  3.4 \\
 2457766.17426 &--12.7 &  3.3 \\
 2457775.00259 &--12.7 &  4.4 \\
 2457775.08258 & --5.6 &  4.4 \\
 2457775.14465 &  14.0 &  4.4 \\
 2457775.17867 &  13.1 &  4.8 \\
 2457775.97223 &   2.9 &  4.4 \\
 2457776.03292 & --0.2 &  4.4 \\
 2457776.07229 & --2.1 &  4.3 \\
 2457776.11589 & --8.8 &  4.8 \\
 2457776.17513 &   7.8 &  4.3 \\
 2457788.03498 &--12.9 &  4.9 \\
 2457788.09157 &--15.1 &  4.8 \\
 2457788.14381 &   4.4 &  5.0 \\
 2457788.96686 & --2.1 &  4.8 \\
 2457789.03347 & --8.5 &  4.6 \\
 2457789.07500 &--19.3 &  4.9 \\
 2457789.12474 & --7.6 &  4.7 \\
 2457789.93510 &--12.3 &  4.7 \\
 2457789.96977 &   0.3 &  5.1 \\
 2457790.02547 &  13.1 &  4.8 \\
 2457790.07589 &   9.3 &  5.1 \\
 2457790.11559 & --0.6 &  5.5 \\
 2457790.94048 &   0.5 &  4.2 \\
 2457790.98777 &   6.0 &  4.2 \\
 2457791.02825 &   2.6 &  4.2 \\
 2457791.06161 &   2.1 &  4.1 \\
 2457791.13065 & --6.2 &  4.1 \\
\midrule
\bottomrule
\multicolumn{3}{l}{$^a$~An additional 6.4~m~s$^{-1}$ was added}\\
\multicolumn{3}{l}{\ \ \ \ in quadrature with these uncertainties}\\
\multicolumn{3}{l}{\ \ \ \ for the RV analyses described in the text.}\\
  \end{tabular}

\end{table}


\begin{table*}
  \centering
  \caption{Radial Velocity Models\label{tab:rvresults}}
  \begin{tabular}{lccccrr}
\toprule
Model              &      trend     &     $K_b$   &     $K_c$   &  dof &$\chi^2$&  BIC\\
                   &   [m~s$^{-1}$~d$^{-1}$]  & [m~s$^{-1}$] & [m~s$^{-1}$]& &      & \\
\midrule
2 planets, trend   &  --0.42$\pm$0.10 & 8.4$\pm$2.1 & 4.8$\pm$2.0  & 50  & 50.0 & 370.2 \\
planet b, trend    & --0.287$\pm$0.084& 6.1$\pm$2.0 &   ---        & 51  & 56.4 & 375.6 \\
planet c, trend    & --0.221$\pm$0.085 &   ---      & 1.9 $\pm$1.7 & 51  & 65.1 & 384.3 \\
no planets, trend  & --0.183$\pm$0.076 &   ---      &  ---         & 52  & 66.1 & 384.4 \\
2 planets, no trend&      ---         & 3.5$\pm$1.8 & 0.3$\pm$1.6 & 51  & 68.2 & 387.4 \\
planet b, no trend &      ---         & 3.5$\pm$1.8 &     ---     & 52  & 68.2 & 386.5 \\
planet c, no trend &      ---         &      ---    &--0.2$\pm$1.6 & 52  & 71.9 & 390.1 \\
no planets, no trend&     ---         &      ---    &     ---     & 53  & 71.9 & 389.1 \\
\bottomrule
\end{tabular}
\end{table*}

\subsubsection{Keck/NIRC2 Adaptive Optics Imaging}

\begin{figure}[bt!]
\begin{center}
\includegraphics[width=3.5in,angle=180]{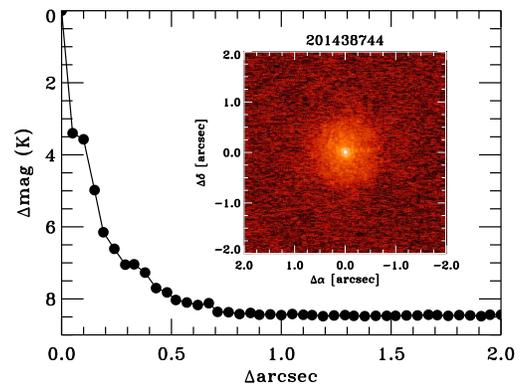}
\caption{\label{fig:keck} We detect no objects near \hd\ in archival
  images or with Keck/NIRC2 adaptive optics, as shown in the image
  (inset) and the resulting $K_s$-band contrast curve. \vspace{0.15in} }
\end{center}
\end{figure}

We obtained Keck/NIRC2 AO imaging of \hd\ on the nights of 2016 Dec
23, 2017 Jan 4, and 2017 Jan 8. Seeing and AO correction were both
poor on the first night, but conditions were good on the second night
and excellent on the third. We therefore use only the third night's
data. We observed using the Br-$\gamma$ filter, a narrow-band $K$-band
alternative that allows us to observe \hd without saturating. We used
the $1024\times1024$ NIRC2 array which has a pixel scale of
9.942~mas~pix$^{-1}$ with the natural guide star system (using \hd\ as
the guide star).  A 3-point dither pattern avoided the noisier lower
left quadrant of the NIRC2 array.  We acquired nine frames with
20 coadds each and a 0.5\,s integration time, and three frames with
40 coadds of 0.5\,s each, for a total of 150\,s of on-source exposure
time.  The data were flat-fielded and sky subtracted and the dither
positions were shifted and coadded into a single final image, shown in
Fig.~\ref{fig:keck}.

The target star was measured with a resolution of 47~mas (FWHM) and we
detect no other stars within the full $10^{\prime\prime}$ field of
view.  We estimate our sensitivity by injecting simulated sources with
S/N=5 into the final combined images at a range of distances from the
central source.  The 5$\sigma$ sensitivities as a function of radius
from the star are shown in Fig.~\ref{fig:keck}. At wider separations,
2MASS J-band imagery shows a possible source 11.2'' north of \hd.
Because the source is not obviously seen in 2MASS H or K, is not in the 2MASS point source catalog, and is not seen in any bands of
UKIDSS, Pan-STARRS, or SDSS, we conclude that it is spurious. We
therefore find no evidence for additional stars within our roughly
40''-diameter photometric aperture.

\subsection{Planet Validation}
Almost all candidates in {\em Kepler}'s multi-planet systems are {\em
  bona fide} planets \citep{lissauer:2011} rather than non-planetary
false positives. Nonetheless, we carry out a full statistical
validation of both transit signals orbiting \hd. As described above,
our HIRES spectrum shows no evidence for secondary spectral lines and
our NIRC2 images show no evidence for secondary stellar
sources. Furthermore, the stellar density inferred from each planet's
light curve fit (assuming a circular orbit; $\rho_{*,circ}$) is
consistent with the stellar density from our \texttt{SpecMatch}
analysis. All these lines of evidence are consistent with a planetary
interpretation of the observed transits.

We therefore follow our previous approach
\citep{schlieder:2016,crossfield:2016} and use \texttt{VESPA}
\citep{morton:2012} along with the NIRC2 contrast constraints and HIRES secondary line
constraints to measure the false positive probability (FPP) of each
transit signal, finding FPP=$4.3\times 10^{-4}$ and $5.1\times
10^{-5}$ for planets b and c, respectively.  Since we see two
transit-like signals, each receives a multiplicity boost that further
reduces the FPPs \citep{sinukoff:2016}.  We therefore conclude that
\hd\ indeed hosts two transiting planets, whose parameters are
summarized in Table~\ref{tab:planet}.

\section{Discussion}
\label{sec:conclusions}
Our analysis indicates two sub-Jovian planets transiting \hd, a bright
($V=8.95$) star, with orbital periods and radii of 9.55\,d and
21.1\,d, and $2.23^{+0.30}_{-0.25}$\,\Rearth\ and
$3.95^{+0.42}_{-0.39}$\,\Rearth, respectively. An RV trend of
\rvtrend\ hints at the presence of a third body at longer orbital
periods.  The stellar parameters are summarized in
Table~\ref{tab:stellar} and the planetary parameters in
Table~\ref{tab:planet}. Below we discuss constraints on the masses,
orbits, and stability of the objects orbiting \hd, and then discuss
future prospects for study of this system.

\subsection{Orbital Dynamics}

Our current RV data are insufficient to measure any planet masses, but
numerous planets with measured masses are known in the
2--4\,\Rearth\ size range \citep{wright:2011}. Examination of the
current mass-radius diagram allows us to estimate masses of 8 and
20\,\Mearth\ for planets b and c, respectively; these estimates are
likely good to roughly a factor of two due to the observed diversity
of envelope fractions among sub-Neptunes
\citep{weiss:2016,wolfgang:2016}. Predictive formulae derived from
planetary mass-radius measurements give results consistent with our
estimate. With these nominal masses, the planets would induce radial
velocity signals with semi-amplitudes of roughly 2.3 and
4.4~m~s$^{-1}$, respectively --- not too far below the system's
RV scatter, indicating that mass measurements will be feasible.  Indeed,
our preliminary RV analysis summarized in Table~\ref{tab:rvresults}
hints that the signals from planets b and c may be detectable and
perhaps larger than predicted in the preceding discussion. Further
observations are needed if we are to adequately sample the two
planets' orbits, disentangle the two planets' signals from other
possible RV noise sources, and robustly measure these planets' masses.

The RV trend we detect indicates that a third body may also orbit
\hd\ at wider separations than planets~b and~c.  Since we do not
detect any curvature, we sample $\lesssim25\%$ of this body's orbit and its
period is $\gtrsim160$\,d.  Following \cite{winn:2009}, for a circular
orbit the minimum mass and semimajor axis of this third object must
satisfy
\begin{equation}
  \frac{M_3 \sin i}{a_3^2}
  \approx (200 \pm 60) M_\oplus\ \mathrm{AU}^{-2} 
\end{equation} Assuming no 
RV curvature,  we know that $a_3 \gtrsim 0.6$~AU, and so the third object
has $M_3 \sin i \gtrsim 45 M_\oplus$. 
Such an object should be at least the size of Neptune and
induce a transit depth of $\gtrsim0.1\%$, which is easily ruled out by
the photometry shown in Fig.~\ref{fig:fits}. If the trend-inducing
object orbits beyond roughly 4.6~AU it would have the mass of a brown
dwarf, and beyond 11.4~AU it must be a star.

Though the two transiting planets are not closely spaced
($a_c/a_b=1.7$), we also evaluate the system's stability. The relevant
length scale for dynamical interactions between planets is the mutual
Hill radius, \RH\ \citep{fabrycky:2012}. Using the planet masses
assumed above, the separation between the two planets 17.4 \RH, much
greater than the minimum separation of $\approx3.5$ necessary for
long-term stability \citep{gladman:1993}. Even if both masses were
twice as large, the separation decreases to only 13.8 \RH. We
therefore conclude that the two planets transiting \hd\ do not violate
the criterion of Hill stability; this conclusion is also consistent
with the observation that many systems discovered by {\em Kepler} and
RV surveys are even more compact. Indeed, there is still plenty of
room: by the above criterion, the system would remain stable even if
another 20\,\Mearth\ warm Neptune orbited between the two transiting
planets.  The 21\,d planet and the third orbiting body are also Hill
stable, having $a_3/a_c > 2.6$ and being separated by $>13\RH$.

Although the system is likely to be dynamically stable, mutual
gravitational perturbations could still cause measurable transit
timing variations (TTVs).  Quantifying the amplitude of any TTVs could
more tightly constrain the masses and orbits than RVs alone
\citep[][]{holman:2010,nesvorny:2013,sinukoff:2017,weiss:2016,weiss:2017}. Assuming
the above planet masses and zero eccentricity, and using the
\texttt{TTVFaster} code of \cite{agol:2016}, we estimate that TTV
amplitudes of up to five minutes could be expected for planet b (whose
mass is presumably lower) and less for planet c. These TTV amplitudes
would tend to increase if either planet has significant eccentricity,
which would be plausible given their small sizes and  orbital
periods. If \hd c's period is not strictly regular, the uncertainty in
its orbital period could be larger than reported in
Table~\ref{tab:planet}. With the entire C10 data set we measure \hd
b's time-of-transit with a precision of only 6.5\,min, so we see no
evidence of TTVs in our {\em K2} data. Nonetheless, precise follow-up
transit photometry might detect such TTVs and would also be sensitive
to additional planets not observed to transit during {\em K2}'s C10
observations. We have planned {\em Spitzer} transit observations of
both planets (GO-13052, PI Werner) to search for TTVs and refine the
orbital parameters of both transiting planets.

\subsection{Follow-up Opportunities}

Because it is bright and because all three bodies orbiting it should
induce measurable radial velocity signals, \hd will be a useful
target. Despite the star's rapid rotation and its radial velocity
scatter  of roughly \rvjitter, our existing observations already suggest
that frequent RV measurements should be able to measure the transiting
planets' masses and constrain their approximate bulk compositions, and
to map the orbit and measure the mass of the third orbiting body.

Another interesting avenue is the mostly-unexplored spin-orbit
alignment of sub-Jovian planets. Though many successful measurements
of the Rossiter-McLaughlin (R-M) effect and of transit tomography have
been made for hot Jupiters, no conclusive R-M measurements have been
made for sub-Neptune-sized planets \citep[but
  see][]{albrecht:2013,lopez-morales:2014,bourrier:2014,barnes:2015}.
Following \cite{gaudi:2007}, the estimated amplitudes of the R-M
effect for planets b and c (assuming spin-orbit alignment) are as much
as 4.2~m~s$^{-1}$ and 12.7~m~s$^{-1}$, respectively, depending on
their (relatively unconstrained) impact parameter. These amplitudes
are not large, but should be measurable. Such measurements are
especially intriguing given the likely presence of the third, long-period
body in the system. Depending on its orbit, long-term interactions
with the inner, transiting planets could have directly impacted their
orbital histories, mutual inclinations, and spin-orbit alignments.

Given the apparent brightness of \hd, the transiting planets could be
useful targets for atmospheric characterization.  The system will be
observable at high S/N by all JWST instruments in most resolution
modes \citep[except the NIRSpec low resolution mode, which will
  saturate;][]{beichman:2015}. Considering their sizes, both planets
likely have considerable volatile content
\citep{marcy:2014,weiss:2014,lopez:2014,rogers:2015planet,wolfgang:2015,wolfgang:2016,dressing:2015b}.
Assuming that these planets have atmospheres dominated by H$_2$/He,
the expected amplitude of spectroscopic signals seen in transit would
be up to 40~ppm in a cloud-free atmosphere (and greater if the planets
are lower-mass than assumed here).  Of those exoplanets studied in
some detail, \hd c is most similar in size and irradiation to
HAT-P-11b (which is slightly larger and more irradiated).  \hd b is
not especially similar to any exoplanet with a well-studied
atmosphere, but is of comparable size to, and lies midway in
irradiation between, HD~97658b and 55~Cnc~e.  Although transmission
spectroscopy suggests that the above planets do not have cloud-free
atmospheres with a low mean molecular weight
\citep{fraine:2014,knutson:2014b}, we expect some sub-Jovian
atmospheres to be amenable to transmission spectroscopy if these
planets' atmospheres are as diverse as those of hot Jupiters
\citep{sing:2016}.  The planets' thermal emission could also be
detected with JWST/MIRI observations: making the gross assumption that
the planets emit as blackbodies, their secondary eclipses have
amplitudes of roughly 20~ppm at 5\,\micron\ and 40--100~ppm at the end
of the MIRI bandpass. Observations of thermal emission would have the
benefit of being relatively unobstructed by any atmospheric aerosols
\citep[e.g.,][]{morley:2015}.

\vspace{0.1in}
Thus the prospects for future characterization are bright for {\em
  K2's} latest multi-planet system. RV spectrographs will quickly
measure the planet masses, determine their spin-orbit alignments, and
transit and eclipse spectroscopy will constrain their atmospheric
makeup. The RV follow-up will also determine the outer body's mass and
orbit, further elucidating the system's architecture.  These detailed
studies will be possible only because they orbit a bright star ---
among the brightest host stars of any {\em K2} systems found to date.
The exciting prospects for future measurements of \hd\ only heighten
our anticipation for {\em TESS}, which we hope will find enough such
systems around even brighter stars to keep the field busy for many
years to come.

\vspace{0.1in} {\bf Note added in review:} While preparing this paper
we became aware of another paper describing the identification of
\hd\ as a planet-hosing system \citep{rodriguez:2017}. We are pleased
that both groups report consistent results despite the fact that no
detailed information was shared prior to  submission of the two papers.

\section*{Acknowledgments}

This work made use of the SIMBAD database (operated at CDS,
Strasbourg, France) and NASA's Astrophysics Data System Bibliographic
Services. This research has made use of the NASA Exoplanet Archive and
the Infrared Science Archive, which are operated by the California
Institute of Technology, under contract with the National Aeronautics
and Space Administration. Portions of this work were performed at the
California Institute of Technology under contract with the National
Aeronautics and Space Administration.  Some of the data presented
herein were obtained at the W.M. Keck Observatory (which is operated
as a scientific partnership among Caltech, UC, and NASA). The authors
wish to recognize and acknowledge the very significant cultural role
and reverence that the summit of Mauna Kea has always had within the
indigenous Hawaiian community.  We are most fortunate to have the
opportunity to conduct observations from this mountain.
I.~J.~M.~C.\ was supported for this work under contract with the Jet
Propulsion Laboratory (JPL) funded by NASA through the Sagan
Fellowship Program executed by the NASA Exoplanet Science Institute.
A.~W.~H.\ acknowledges support for this K2 work from a NASA
Astrophysics Data Analysis Program grant, support from the K2 Guest
Observer Program, and a NASA Key Strategic Mission Support Project.
LMW acknowledges support from the Trottier family.

{\it Facility:} \facility{Kepler}, \facility{K2},  \facility{Keck-II (NIRC2)},  \facility{Keck-II (HIRES)}


\end{document}